\documentclass[useAMS,usenatbib,usegraphicx]{mn2e}
\title[Predictions for CARMA and ALMA]
 {Observing Turbulent Fragmentation in Simulations: \\
 Predictions for CARMA and ALMA}
\author[Offner et al.] 
 {Stella S.~R.~Offner,$^1$\thanks{E-mail: soffner@cfa.harvard.edu} John Capodilupo,$^2$ Scott Schnee$^3$ and Alyssa A.~Goodman$^1$
\\
 $^1$Harvard-Smithsonian Center for Astrophysics,
     Cambridge, MA 02138 \\
 $^2$Harvard University, Cambridge, MA 02138 \\
 $^3$NRAO, Charlottesville, VA 22903}

\begin{document}

%\date{Accepted 1988 December 15. Received 1988 December 14; in original form 1988 October 11}

\pagerange{\pageref{firstpage}--\pageref{lastpage}} \pubyear{2011}

\maketitle

\label{firstpage}

\begin{abstract}

Determining the initial stellar multiplicity is a challenging problem 
since protostars are faint and deeply embedded at early times;
once formed, multiple protostellar systems may significantly
dynamically evolve before they are
optically revealed. Interferometers such as CARMA and ALMA
make it possible to probe the scales at which turbulent fragmentation
occurs in dust continuum emission, potentially constraining early stellar multiplicity. In this Letter we
present synthetic observations of starless and protostellar cores
undergoing fragmentation on scales of a few thousand AU to produce wide binary systems. We show that
interferometric observations of starless cores by CARMA should be
predominantly featureless at early stages, although wide protostellar
companions should be %readily 
apparent. The enhanced capabilities of ALMA
improve the detection of core morphology so that it
may be possible to detect substructure at earlier times. In either
case, spatial filtering from 
interferometry reduces the observed core substructure and often eradicates traces of existing
filamentary morphology on scales down to 0.025 pc. However, some missing structure may be
recaptured by combining data from the ALMA full science and Atacama compact arrays.  
 \end{abstract}


\section{Introduction}

While the initial mass function of stars is well measured, the
initial stellar multiplicity remains largely
unconstrained \citep{duchene07}. Determining multiplicity among field
stars well after
the formation process has ended is itself a
challenging problem: close binaries are often unresolvable,
gravitational boundedness is difficult to confirm,  
and background stars are ever-present
interlopers \citep{raghavan10}.
During the formation and subsequent evolution, dynamical
interactions influence initial companion separations and decrease multiplicity. 

Observing the initial protostellar multiplicity in situ is challenging
for similar reasons with the added complication that young protostars
are dim and heavily obscured by dust and gas. However, recent
interferometric instruments such as the Combined Array for Research in
Millimeter-wave Astronomy (CARMA) and the Atacama Large
Millimeter/submillimeter 
Array (ALMA) are revolutionizing the ability to probe the earliest
stages of cores on scales of a few arcseconds or less. This presents
an opportunity to observe core substructure, protostellar disks, and young companions.

There are two main theories of binary star formation. In the disk
fragmentation scenario, massive protostellar accretion disks become
Toomre unstable and fragment into one or more close companions
\citep{ARS89, bonnell94}. In the
turbulent fragmentation scenario, turbulent perturbations within a single
prestellar core or filament 
individually collapse to form separate stars with wide separations
\citep{goodwin04, fisher04, goodwin07}. 
Numerical and analytic arguments indicate that disk instability should
 be quite common in high-mass star formation 
\citep{kratter06, kratter08}. However radiation feedback significantly
reduces the fragmentation of disks around low-mass stars, resulting in few low-mass
multiple star systems \citep{cai08, Offner09, bate09}. The
multiple systems
that do form are the result of turbulent fragmentation and have initial
separations of $\sim$1000 AU \citep{offner10}.

%Plot 1
\begin{figure*}
%\epsscale{1.0}
%\setlength\fboxsep{0pt}
%\fbox{
\includegraphics[height=6.2cm, width=14cm]{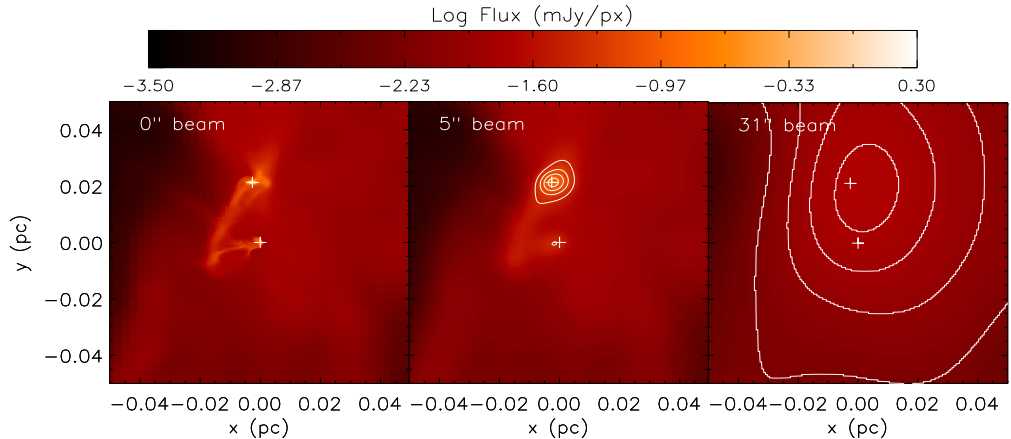}
 \caption{Single dish observation of 1.1 mm flux in mJy per pixel (0.13'') of a binary protostellar system
  formed via turbulent fragmentation, where the system is placed at
  250 pc.  Left: Perfect resolution simulation
  data. Center: Image convolved with a 5'' beam. Right: Image convolved with a 31'' beam. The
  image extent is 80'' across at a distance of 250 pc. Crosses mark the
  protostar positions.  Contours indicate 30\%, 50\%, 70\% and 90\% of the
  image maximum. }
 \label{singledish}
\end{figure*}

The enhanced capabilities and sub-arcsecond resolution of ALMA will
make it possible to test both
scenarios of binary formation and map protostellar accretion disks
down to
AU size scales.
Several authors have used interferometric observations to look
  for young companions of Class 0 objects (e.g., \citealt{looney00,jorgensen07,
    maury10}). However, they arrive at somewhat different conclusions: \citet{looney00} find that
  all their targeted embedded objects have companions, while
  \citet{maury10} find only one tentative companion in a sample of
  five Class 0 sources. Likewise, \citet{jorgensen07} identified only
  one candidate previously unknown companion.
Using continuum emission, it is possible in principle to detect
companions forming as a result of turbulent fragmentation at even earlier times.
%Using CARMA, it is currently possible to probe for 
%fragmentation of cores in nearby star forming regions. 
For example, \citet{schnee10} use CARMA to observe 3mm continuum emission from 11 starless 
cores in the Perseus molecular cloud at 5" resolution.
They
found that the cores had no conspicuous substructure. This could be
due to several possibilities. The cores may be too young or may never go on to form an
individual star much less a wide multiple system. Turbulent
fragmentation may be very uncommon. Or turbulent fragmentation may be
ongoing in the cores, but beyond the observable limits of CARMA. We
investigate the third possibility here.

%\citet{Offner10} find a number of instances of wide binary formation,
%where fragmentation initially takes place on scales of a few thousand
%AU. Such scales are intermediate between scales of disks (few 100 AU)
%and the scale of the host core or filament (0.05-0.1 pc). 
If cores, such as those investigated by \citet{schnee10}, actually
contain young protostars, 
it might be possible to
observe fragmentation at shorter wavelengths. Indeed,  
a number of very cold cores thought to be
starless have since been found to contain protostars \citep{enoch10, dunham11,
  pineda11}. However, in the case of L1451, the presence of a 
protostar, potentially still in the ``first core'' gas stage,  was
ultimately identified by 
outflow activity rather than by thermal emission. This
reinforces the point that source identification is
challenging, and it requires instruments with high sensitivity and
resolution  
% in order
to characterize core structure and identify young companions. 

In this Letter, we use the CASA software package to synthetically
observe binaries forming due to turbulent fragmentation in
the numerical simulations of \citet{Offner09}.
By following the evolution of such pairs beginning in the
prestellar core stage, we can make predictions about the feasibility of
observing such fragmentation in dust continuum and constraining stellar multiplicity at
the earliest stage of star formation. 

%In section 2, we describe our
%simulation and methods. In section 3, we present synthetic
%single-dish observations and interferometric CARMA and ALMA
%observations. We summarize our conclusions in section 4.

% call_fig0 -> multi_amr_fig0

\begin{figure*}
\includegraphics[height=9.0cm, width=17cm]{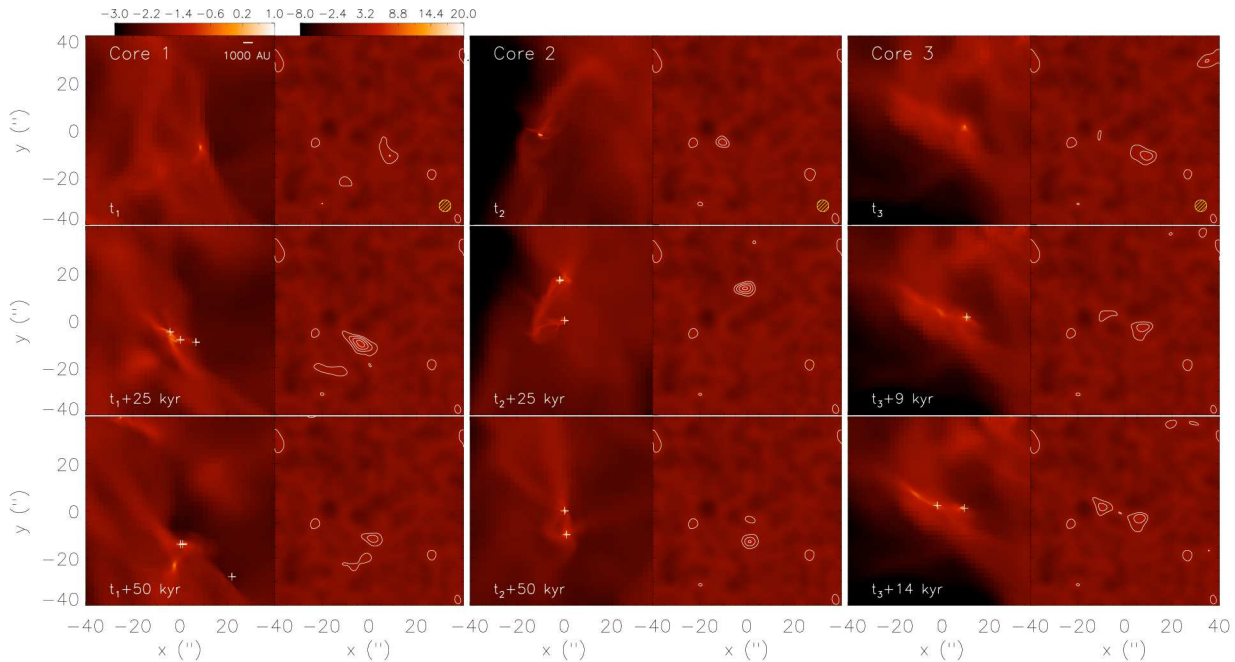}
\caption{Comparison of simulations and CARMA synthetic observation
  for fragmenting cores at three different times. Left: Simulation data
  with perfect resolution in 1.1mm flux (Log mJy beam$^{-1}$). Color scale
  is the same as in Figure 1. Right: CARMA
  observation of each core at 3mm emission including Gaussian
  $\sigma_{\rm S}$ noise assuming a distance of 250 pc in mJy beam$^{-1}$. Contours indicate 30\%, 50\%, 70\% and 90\% of the
  maximum flux. The CARMA beamsize is indicated by the hatched ovals.
\label{carma} }
\end{figure*}

\section{Numerical Simulations and Methods}

The molecular cloud simulations we observe in this letter are those presented in
\citet{Offner09}. Since \citet{Offner09} fully describe the calculations,
we include only a brief overview here. The calculations are performed
using the ORION adaptive mesh refinement code including driven
large-scale turbulence, self-gravity, and flux-limited diffusion
radiative transfer. Forming stars are modeled by sink particles with a sub-grid stellar evolution
model. These are inserted in regions of the flow that exceed the maximum
grid resolution. In practice, since protostellar winds are not included, these particles
give an upper limit on the stellar mass.  The cloud domain size is 0.65 pc across with a minimum cell
size of 32 AU.

In our analysis, we focus on several typical systems out of the
approximately 10 close pairs that form over the course of the
simulation. Since the systems form self-consistently from the
turbulent gas, the
fragmentation history and stellar masses are not predetermined.
 
We use the ``simdata" task in the Common
Astronomy Software Applications (CASA)
package\footnote{http://casa.nrao.edu} 
to produce synthetic interferometric observations of the starless and
protostellar systems.  Using simdata, the model cloud was placed at
the RA and Dec of the Perseus molecular cloud.   
For the CASA input, we require maps of the cores in
units of flux. Converting between simulated column density and
continuum flux is straight-forward for optically thin emission. Assuming a constant dust
temperature, $T_D$, the flux at a given frequency, $\nu$, can related
to the column density by:
%This is from Enoch et al. 2007, eq 3
\begin{equation}
S_{\nu} = \Sigma B_{\nu}(T_D) \kappa_{\lambda} \Omega_b, \label{dust}
\end{equation}
where $B_{\nu}(T_D)$ is the Planck function evaluated at the dust
temperature, $\kappa_{\lambda}$ is the dust opacity, $\Omega_b$ is the
beam solid angle and $\Sigma$ is the
gas column density per pixel (e.g., \citealt{enoch06}). The opacities for 1.1 mm and 3 mm are
$\kappa_{1.1} = 0.0114$ cm$^{2}$ g$^{-1}$ and $\kappa_{3} = 0.00169$
cm$^{2}$ g$^{-1}$ \citep{ossen94}. The simulations assume that the dust and gas
are well-coupled, which is a reasonable approximation for
densities $> 10^4$ cm$^{-3}$ \citep{goldsmith78}. Thus, we adopt $T_{D}= 10$ K, the
simulation gas temperature, which is also a lower limit for the dust
temperature in Perseus \citep{schnee09}. 
Once
protostars form, they heat their environment to
 mean temperatures of 15-20 K and result in higher emission.
 
Producing synthetic single dish observations of the
simulations is a straight-forward
application of equation \ref{dust}. Finite resolution can then be imposed by
convolving the flux map with a circular Gaussian beam.
%In these cases, we could directly use the gas temperatures from the
%simulations. However, these are probably over-estimates of the gas
%temperature, and the cores appear bright enough at 10 K to be seen.

\section{Synthetic Observations}

For the purpose of comparison, we first present synthetic single dish 
observations of the fragmenting cores. We then produce
interferometric observations mimicing the specifications of CARMA and
ALMA. Finally, we investigate the influence of noise
and distance on structure detection.

\subsection{Single Dish Observations}

Observations that map out entire molecular clouds and identify star-forming cores often use single dish continuum data.
However, with beam resolutions of tens
of arcseconds, resolving core substructure in even nearby clouds is
impossible.  
Figure \ref{singledish} shows a protostellar system forming
in a dense core with 0'' (i.e., perfect), 5'' and 31''
resolution. The two peaks are distinct with 5'' resolution, although
the fainter peak may not be apparent depending upon sensitivity and
noise levels. At 31'', a resolution comparable to SCUBA (850 $\mu$m)
and Bolocam (1.1 mm), only a single peak is apparent.

Interferometers, such as CARMA, are able to achieve this 5'' scale, but
can only probe a fixed window of scales. Since larger scale
information is resolved out, target fluxes may also be reduced by 90\% or
more relative to single dish observations.  
Figure \ref{singledish} 
illustrates that cores forming wide protobinary companions {\it may} be
observable with current interferometer technology.

%If the observed sources shown
%warmed the gas sufficiently and were not too deeply embedded, it might
%be possible to identify the two in shorter wavelength
%emission. However, a number of cores that were very cold and thought to be
%starless have since been found to contain protostars \citep{enoch10, dunham11,
%  pineda11}. In the case of L1451, the location of the
%forming star, potentially still in the ``first core'' gas stage,  was ultimately identified by the
%presence of an outflow rather than thermal emission. This
%reinforces the point that source identification is
%challenging, and it requires instruments that probe smaller scales to
%characterize core structure and potential young companions. 
%make_fig2.pro

\subsection{CARMA Observations}

We produce synthetic observations to compare with the CARMA observations of 
\citet{schnee10}. In order to make a truly similar comparison, we reproduce their
  observing procedures and conditions as closely as possible.  Our observations have a 2.8 GHz  % Is this (current bw
				% of CARMA) right or is this 2.8 (old
				% Schnee bw)?
%The main thing that the bandwidth is important for is for calculating the noise in the map, which goes down when you average over larger bandwidths.  Since we added the noise after the fact, and not based on the bandwidth in the simulation, we would get the same maps with an "inwidth" of 2, 2.8, or 4 GHz.  We can say whatever we like here, but since the noise we added was the appropriate amount for the 2.8 GHz bandwidth, let's just say we did that.
bandwidth centered at 102 GHz, and we adopt the CARMA D-array
configuration and an integration time of 100 seconds with a total time
of 8 hours. Unless otherwise stated,
we assume the systems are 250 pc away: the distance of the Perseus
molecular cloud. The pixel size of the simulated map is 0.13'', and the synthesized beam
is ~5''.
We add synthetic
Gaussian noise with $\sigma_{\rm S} = 0.7 $ mJy beam$^{-1}$, comparable to
that of the \citet{schnee10} observations.

%make_fig3.pro
\begin{figure}
\includegraphics[width=8.5cm]{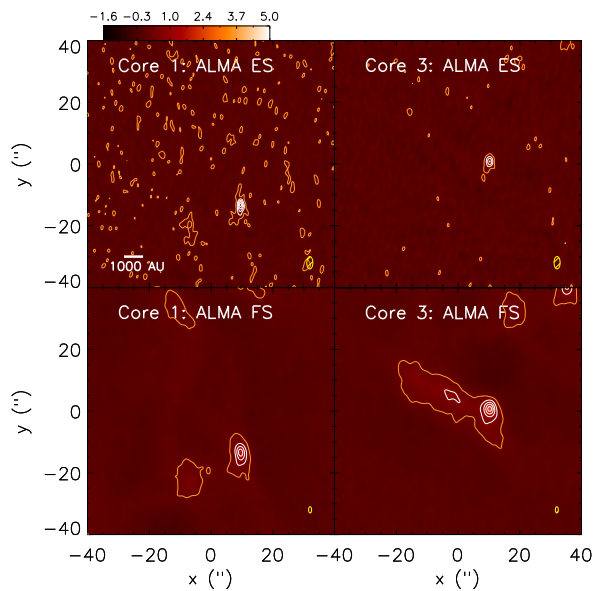}
\caption{Comparison of early science (ES) and full science (FS) ALMA
  configurations for Core 1 (left) at $t_1$ and Core 3 (right)
  at $t_3$ assuming a distance of 250 pc. Color scale indicates flux in
  mJy beam$^{-1}$. The top two panels are observed with 
  the early science (ES) extended configuration; the bottom two panels are
  observed in one example full science (FS) configuration. White contours indicate 30\%, 50\%, 70\% and 90\% of the
  maximum flux. Orange contours indicate 10\% of the maximum flux. The ALMA beamsize is indicated by the hatched ovals.
\label{alma} }
\vspace{-0.4cm}
\end{figure}

%make_fig4.pro
\begin{figure}
\includegraphics[height=11cm, width=7cm]{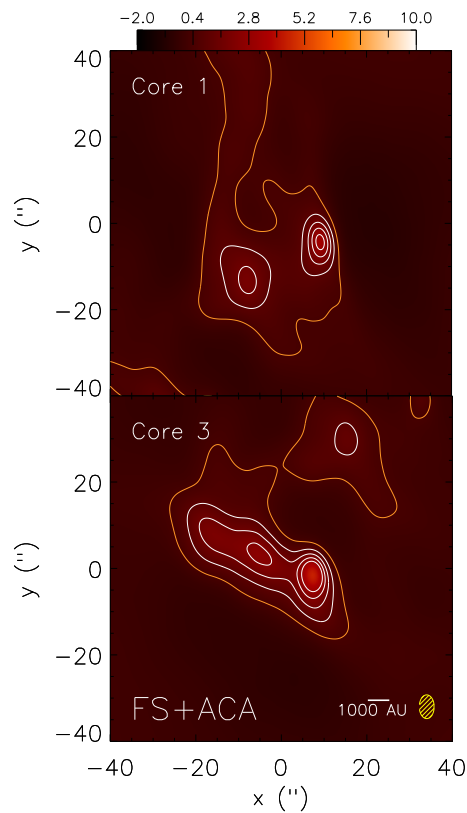}
\caption{Synthetic observation combining data from ALMA FS with ALMA compact
  array for Core 1 (top) at $t_1$ and Core 3 (bottom) at $t_3$.
 Contours indicate fluxes with values 10\% (orange), 30\%, 50\%, 70\% and 90\% of the
  maximum. The effective beamsize is indicated by the hatched oval. The
  observations assume 100s integration time and two hours of observing
  with each configuration.
\label{aca} }
\end{figure}

Figure \ref{carma} shows synthetic CARMA observations of three separate
cores undergoing turbulent fragmentation at three different times. The
simulation data seen in 1.1 mm emission is plotted for
comparison.
At the first time all
three cores are starless. However, each shows some evidence of
collapse and
fragmentation at perfect resolution. The second row illustrates that fragmentation proceeds on the
order of 10-20 kyr, which is a relatively short time compared to the
typical $\sim$100 kyr dynamical time of cores. If the core exists in a
quasi-steady state for some time before undergoing collapse (e.g., \citealt{broderick}), then the
likelihood of catching any particular starless core in the act of
fragmenting may be small. 
Such short timescales for
observing close companions are consistent with those found by
\citet{stamatellos11}, who synthetically observed massive fragmenting disks.
The probability is reduced further since not
all cores (may) experience fragmentation on $\sim$ 1000 AU scales.

At the earliest times, interferometric observations do not clearly
show fragmentation. The second fragment in Core 1 becomes more
apparent over time, but is nearly invisible in the starless
phase. However, the low level detections do appear similar in size and separation to some
of the lower flux contours apparent in maps of Perbo45 and Perbo58 by \citet{schnee10}.
In contrast, all of the filamentary structure in the starless Cores 2 and 3
is resolved out. 
%Part of the fragmenting filament of Core 3 is delineated by the lowest contour,
%which indicates that fragmentation or core structure may only be observed
%within $\sim$10 kyr of protostar formation. 
These results are consistent with the core observations of
\citet{schnee10}, who find that most starless cores that have
  bright 1.1mm emission in single dish maps are undetected in 3mm
  interferometric maps. This also confirms 
that additional filamentary structure may be removed by the
interferometric technique.

The synthetic observations show that at later times the 
  protostars and companions become brighter. Due to the factor of 10
  difference in opacity, the over-densities will be significantly
  brighter in 1.1mm, which will increase the signal to noise.
%individual
%stars appear very clearly and are reasonably well resolved. 
This
suggests that widely separated protostellar companions should be
relatively apparent at high resolution (e.g., \citealt{merrill10}). 
However, the length of the window in which fragmentation occurs is still problematic if secondaries
migrate to shorter separations or are unbound on short timescales. 

%make_fig5.pro
\begin{figure*}
\includegraphics[height=6.5cm, width=15cm]{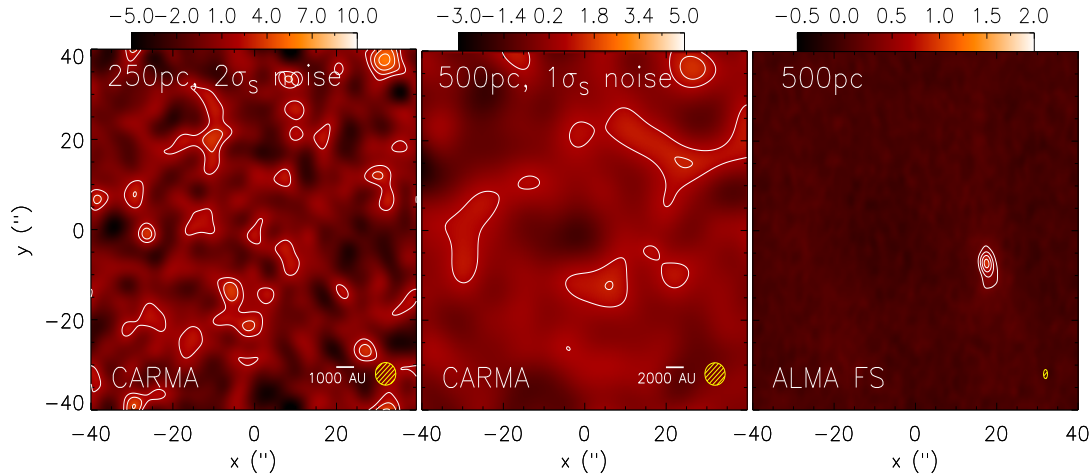}
\caption{Left: Simulated 3mm CARMA observation of Core 1 at
  $t_1$ observed at a distance of 250 pc
  with 2$\sigma_{\rm S}$ noise. Center: Simulated 3mm CARMA observation of Core 1 observed at a distance of 500 pc
  with 1$\sigma$ noise.  Right: Simulated 3mm full science ALMA
  observation of Core 1 observed at a distance of 500 pc.  
  Color scale shows flux in mJy beam$^{-1}$. 
 Contours indicate fluxes with values 30\%, 50\%, 70\% and 90\% of the
  maximum. The beamsize is indicated by the hatched ovals.
\label{noise} }
\end{figure*}

\subsection{ALMA Predictions}

ALMA, which will have 66 reconfigurable antennas, sub-arcsecond angular resolution and sensitivity
to wavelengths from 3mm to 300 $\mu$m,  will 
significantly expand observational capabilities. Although currently only
partially completed, the ALMA early science program allows
reduced observations with 16 12m operational antennas. Figure
\ref{alma} shows simulated observations of Core 1 and Core 3 comparing
the ALMA early science Cycle 0 extended
configuration and the ALMA full science configuration 10.  We adopt
configuration 10 in the ALMA full science library in simdata since it 
is a representative intermediate configuration. 

%PWV = 2.8 was chosen because that's what the ALMA sensitivity calculator uses to estimate the noise for observations made around 100 GHz.  If the weather were much better than that, the telescope would probably observe at a higher frequency.  If the weather were much worse, it probably wouldn't observe (usefully) at all.  I chose the CLEAN threshold to be ~3 times the rms noise I expected in the maps, based on the ALMA sensitivity calculator.  Cleaning to about 3 sigma is standard practice, with the idea that going to a lower threshold just moves noise around the map without finding any additional "real" signal.
Each observation
assumes a 100 second integration time and a total time of 2 hours. 
We add thermal noise to the simulated ALMA images using simdata,
assuming that the precipitable water vapor (PWV) during the
observations is 2.8mm (appropriate for Band 3 ALMA observations) and a
bandwidth of 8 GHz.  ALMA maps were cleaned until a threshold of 3
times the rms thermal noise in the map was reached.  The pixel size,
as for the CARMA simulations, is 0.13'', and the early science and
full science synthesized
beam sizes are $\sim$3'' and 1.5'', respectively.
%It's the 10th most compact configuration out of 28 example ALMA configurations, so it seemed to be a representative (i.e. intermediate) configuration.  The beam size is about 2-3 times smaller than CARMA (and twice as high as the ALMA ES compact configuration), so it seemed like a big, but still roughly comparable, improvement in resolution.  I can use another configuration if you prefer.

Figure \ref{alma} illustrates that ALMA better resolves the primary
peak morphology. A secondary peak in Core 1 is marginally shown by contours with 10\% of
the maximum flux. However, the filamentary structure of Core 3 is not visible.
The early science configuration sensitivity to
substructure is otherwise similar to that of CARMA with four times
the observing time. The ALMA full
science configuration detects the Core 3 filament with some suggestion
of further fragmentation. This highlights the difficulty of imaging 
structure and fragmentation even with ALMA's superior resolution.

It is possible to increase the range of recoverable
spatial scales by combining data from antenna
configurations with different baselines.
The size of the ALMA 12-meter antennas prohibit them from being placed
  within 15 meters of one another, which limits the maximum resolvable
  spatial scale. To permit a greater range of observations, 
the completed ALMA site will include a
  second smaller array, the Atacama Compact Array (ACA), comprised of
  four 12-meter and 12 7-meter antennas. 
Here, we combine the higher spatial
resolution synthetic data of the intermediate FS main array configuration with
synthetic data from the 7-meter ACA antennas. Figure \ref{aca} shows a simulated observation
of two cores where the visibilities from FS 
and ACA have been added and
then deconvolved using the CASA CLEAN subroutine. The appearance of filamentary structure
and core substructure is significantly improved compared to the
single configuration data in Figures \ref{carma} and \ref{alma}.

\subsection{Noise and Resolution Limitations}

Object distance and observation sensitivity are both critical to
mapping  
core structure. Figure \ref{noise} shows Core 1 with
increased noise and source distance. 
%While the lowest contour
%indicates a secondary peak
At our fiducial noise and resolution (see Figures
\ref{carma} and
\ref{alma}) both increased noise and distance eliminate any
detection with CARMA. At a distance of 500 pc, the secondary peak is
also undetected with the ALMA full science configuration. However, the primary
peak is reliably detected. This implies
that fragmentation at later times, such as that illustrated in Core 1 at 50 kyr and
Core 3 at $\ge 9$ kyr, would be readily observable.

\section{Conclusions}

In this Letter we
produce synthetic observations of fragmenting starless and protostellar cores.
We show that
interferometric observations of starless cores by CARMA should be
predominantly featureless at early stages. In fact, structure may be
apparent only within a short period, $\sim 10$kyr, of the formation of a protostar. 
This may account for some of the
apparent lack of substructure in starless cores noted by
\citet{schnee10}. We find that wide protostellar
companions with separations of $\sim 1000$AU should be
detectable. The confidence of the secondary detection depends upon
  the source age, resolution, and signal-to-noise, which may partially explain the
  differing results of \citet{looney00}, \citet{jorgensen07}, and \citet{maury10}. ALMA's
enhanced capabilities 
improve the detection of core morphology, so that it
may be possible to detect substructure at earlier times. Filamentary
structure is more difficult to detect than peakiness, and
interferometry, especially at high resolution, significantly reduces the
presence and extent of filaments. However, we find that it is possible
to recover missing structure by combining ALMA Full Science data with
data from the Atacama Compact Array.

\section*{Acknowledgments}

This research has
been supported by the NSF through grants AST-0901055 (SSRO) and
AST-0908159 (AAG) and the Harvard College Program for Research in Science and Engineering (JC).

%\bibliography{OKbib.bib}
%\bibliographystyle{mn2e}

\end{document}